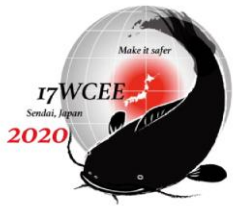

# Deep Bayesian U-Nets for Efficient, Robust and Reliable Post-Disaster Damage Localization

X. Liang [(1)], S.O. Sajedi [(2)]

[(1)] *Assistant Professor of Research, Department of Civil, Structural and Environmental Engineering, University at Buffalo, the State University of New York, Buffalo, NY, United States, liangx@buffalo.edu*
[(2)] *Ph.D. Candidate, Department of Civil, Structural and Environmental Engineering, University at Buffalo, the State University of New York, Buffalo, NY, United States, ssajedi@buffalo.edu*

## *Abstract*

Post-disaster inspections are critical to emergency management after earthquakes. The availability of data on the condition of civil infrastructure immediately after an earthquake is of great importance for emergency management. Stakeholders require this information to take effective actions and to better recover from the disaster. The data-driven structural health monitoring (SHM) has shown great promises to achieve this goal in near real-time. There have been several proposals to automate the inspection process from different sources of input using deep learning (e.g., visual inspections and vibration data). The existing models in the literature only provide a final prediction output, while the risks of utilizing such models for safety-critical assessments should not be ignored. This paper is dedicated to developing deep Bayesian U-Nets where the uncertainty of predictions is a second output of the model, which is made possible through Monte Carlo dropout sampling in test time. Based on a grid-like data structure, the concept of semantic damage segmentation (SDS) is revisited. Compared to image segmentation, it is shown that a much higher level of precision is necessary for damage diagnosis. To validate and test the proposed framework, a benchmark dataset, 10,800 nonlinear response history analyses on a 10-story-10-bay 2D reinforced concrete moment frame, is utilized. The location and probability ratio of the dropout operator as well as the Monte Carlo sampling size are investigated by a thorough parametric study, showing that how these hyperparameters affect the overall robustness. Moreover, four different strategies in adjusting the observation weights and modifying the decision rule are considered in all case studies. Compared with the benchmark SDS model, Bayesian models exhibit superior robustness with enhanced global and mean class accuracies. Finally, the model's uncertainty output is studied by monitoring the softmax class variance of different predictions. It is shown that class variance correlates well with locations where the model makes mistakes. This output can be used in combination with the prediction results to increase the reliability of this data-driven framework in structural inspections.

*Keywords: SHM, Damage Diagnosis; Post-disaster inspections; Deep learning; Bayesian Inference*

## 1. Introduction

Post-disaster condition assessment is an integral part of emergency management after an earthquake. A decision-maker requires reliable information to minimize the downtime of civil infrastructure and to accelerate recovery. Conventionally, human inspectors have been in charge of monitoring the structural conditions where the evaluations are mainly through visual inspections. Past studies have shown that human-based condition assessments are subjective and thus prone to errors [1]. Moreover, to minimize the negative social and economic consequence of an earthquake, the post-disaster inspections of civil infrastructure has to be done in a very time-efficient manner. Having teams of trained individuals to identify damage in a significant number of structures is not only impractical but also may raise safety concerns for the inspectors.

By considering those limitations, the need for automated structural health monitoring (SHM) is highlighted. The recent advances in artificial intelligence (AI) have had a significant impact on different fields of engineering, including SHM. Deep learning algorithms have proven to be robust tools that can learn highly sophisticated and nonlinear mappings from real-world observations. Image-based inspection and monitoring (e.g., using drones [2, 3]) benefited significantly from the recent breakthroughs in computer vision. For example, convolutional neural networks prove to be very robust in identifying cracks from raw images without a prior need for image processing or human interventions (e.g., [4-6]). Several studies investigated the application of deep learning to localize/identify various types of visible structural defects from raw images [7-9]. Structural object detection is another area where such information is used in the automatic guidance of drones used in bridge inspections [10-12]. While having its own merits, visual condition assessments only describe the surface damage. Images commonly provide a qualitative description of the restricted scene that is captured by a camera. It is challenging to visually assess the global condition and safety of a civil infrastructure unless it is near collapse, when the damage is obvious. Moreover, visual access to critical structural components is commonly restricted because they are not often exposed.

Another approach in SHM is to monitor the changes in the structural dynamic behavior, which is reflected in the vibration data. The severity and location(s) of the damage are associated with specific changes in properties such as stiffness and damping. An advantage of vibration-based SHM is that obtaining acceleration records from buildings and bridges is relatively easier than data from other types of instrumentation (e.g., strains). This is mainly because recording acceleration response of an existing building is less challenging than recording deformation and forces. System identification methods have been used for damage diagnosis using acceleration records (e.g., [13]). However, the application of such techniques requires expert supervision for implementation in near real-time and, therefore, can be time-consuming.

A more recent and trending area of SHM research investigates the application of machine learning algorithms [14]. Data-driven models can potentially estimate an input-output mapping between the vibration data and damage. This mapping could be highly complex due to the nonlinearity in the structural response, subjected to earthquake excitation and various sources of environmental noise. Support vector machines [15-17], clustering [18], and neural network-based applications [19, 20] are examples where classical machine learning algorithms are proposed for damage diagnosis in specific case studies. A potential issue with these methods is that their real-time implementation is challenging and the corresponding cost of computation is high when dealing with large input sizes. Given the success of convolutional neural networks (CNNs) in the computer vision, more recent efforts have adopted CNNs in vibration-based SHM [21-23]. Therefore, the acceleration records from multiple sensors serve as the initial input in the process of automatic feature extraction. In our latest work, we extended the idea of pixel-wise image classification to vibration-based SHM [24]. Therefore, a semantic damage segmentation (SDS) framework is developed where all the nodes in a building (similar to pixels in an image) are monitored for the presence of damage. [24] verified the potential robustness of the SDS concept for the damage diagnosis by thousands of nonlinear time history analysis simulations.

Based on the brief literature review, AI portrays a bright future for automated SHM. However, the integration of such data-driven models in the SHM industry still faces challenges. One of the most critical concerns in this regard is the fact that the use of a data-driven classifier is always associated with the risks of making mistakes. Unlike the computer vision applications where a deep learning architecture is trained and tested on millions of images from different environments, the validation of the existing data-driven SHM models is commonly limited to a few case studies. Lack of sufficient quality damage data will potentially increase the chances of error in a data-driven model. These models are used to directly monitor the safety of civil infrastructures that can affect populated areas. Therefore, the consequences of mistakes can be fatal. The level of prediction accuracy (robustness) can vary in data-driven models depending on the algorithm and the dataset used in training. Nonetheless, the chances of misclassification are inevitable.

In this paper, we develop the Bayesian U-Nets based on the concept of SDS, where predictions are accompanied by a measure of the model's uncertainty. Knowledge about the prediction uncertainty is essential in a damage diagnosis model because it can trigger human intervention. As a result, a human operator can decide whether or not further inspections are necessary for a reliable estimate of damage.

## 2. Semantic Damage Segmentation

As briefly mentioned in the literature review, CNNs have shown great success in processing raw image data into valuable information that can be used for structural inspections. Their robustness is combined with computational efficiency, where the network's learnable parameters (feature maps) are shared in the sliding kernels. Semantic scene segmentation is an example use of CNNs where all the pixels in an image are classified with a corresponding label.

The difference between vision scene and SHM damage segmentation is that for the vision task, roughly identifying the pixels that describe an object may be sufficient. However, in damage segmentation, every single structural node (equivalent to pixels in an image) should be correctly identified to judge the safety of a building. For example, the collapse in a moment frame can be initiated by the formation of a soft story mechanism where the majority of structural elements remain locally intact. This issue is better explained through an example shown in Fig. 1. The pixel-wise prediction of the road scene, though not perfect, is probably sufficient for an autonomous vehicle to avoid an accident. The geometrical equivalents of pixels in damage segmentations are the grid nodes (i.e., blue squares in Fig. 1.b [24]). Missing damage in a grid node can potentially have fatal outcomes. Therefore, the correct prediction of all grid nodes is essential for reliable damage diagnosis. One should also note that, commonly, there is a substantial difference between the resolution of an image and the grid environment.

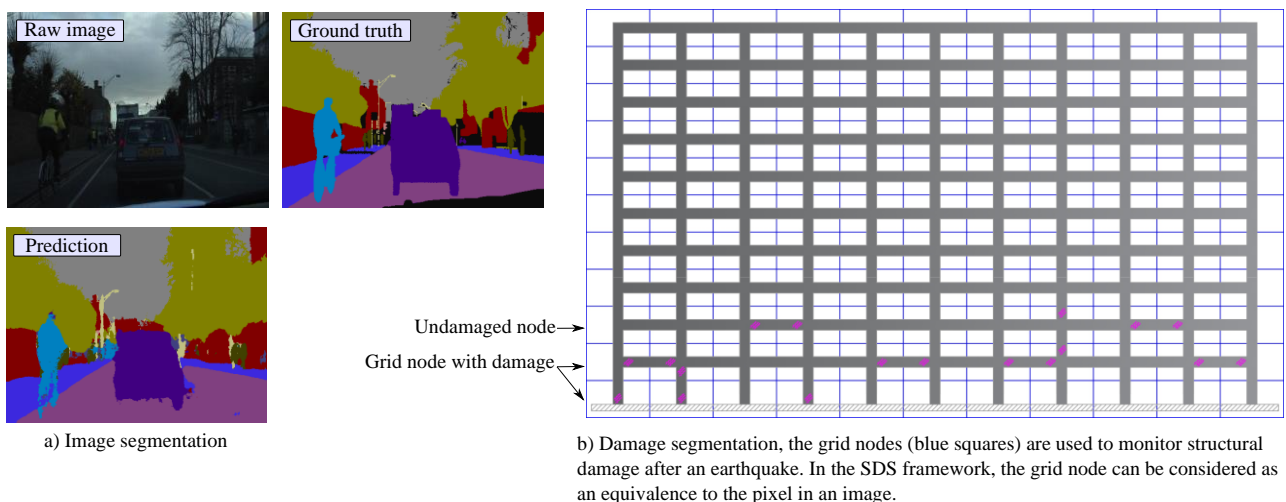

a) Image segmentation

b) Damage segmentation, the grid nodes (blue squares) are used to monitor structural damage after an earthquake. In the SDS framework, the grid node can be considered as an equivalence to the pixel in an image.

Fig. 1 – Comparison of semantic segmentation between computer vision and vibration-based SHM. (the images in (a) are from [25]).

## 3. Bayesian U-Nets

Uncertainty quantification in this study is performed by fusing Bayesian inference in the U-Nets [26]. This deep learning architecture was first proposed for biomedical image segmentation and is known for its robustness and computational efficiency. We are inspired by the architecture of U-Net and modify it for SDS. The deep learning architecture is made of two paths connected by a bottleneck. The contracting part is made of several computational modules that include a sequence of convolution, batch normalization, ReLu activation, and max-pooling layers. In the expansive (upsampling) path, there is a conjugate computational block for each module in the contracting path. Each expansive module is respectively comprised of transposed convolution, concatenation, convolution, batch normalization, and ReLu activation. Note that the concatenation layers combine the feature maps extracted from the convolution layer in the contracting path with the corresponding transposed convolution in the expansive path. The architecture is concluded by 1×1 convolutions and a softmax activation for class probabilities.

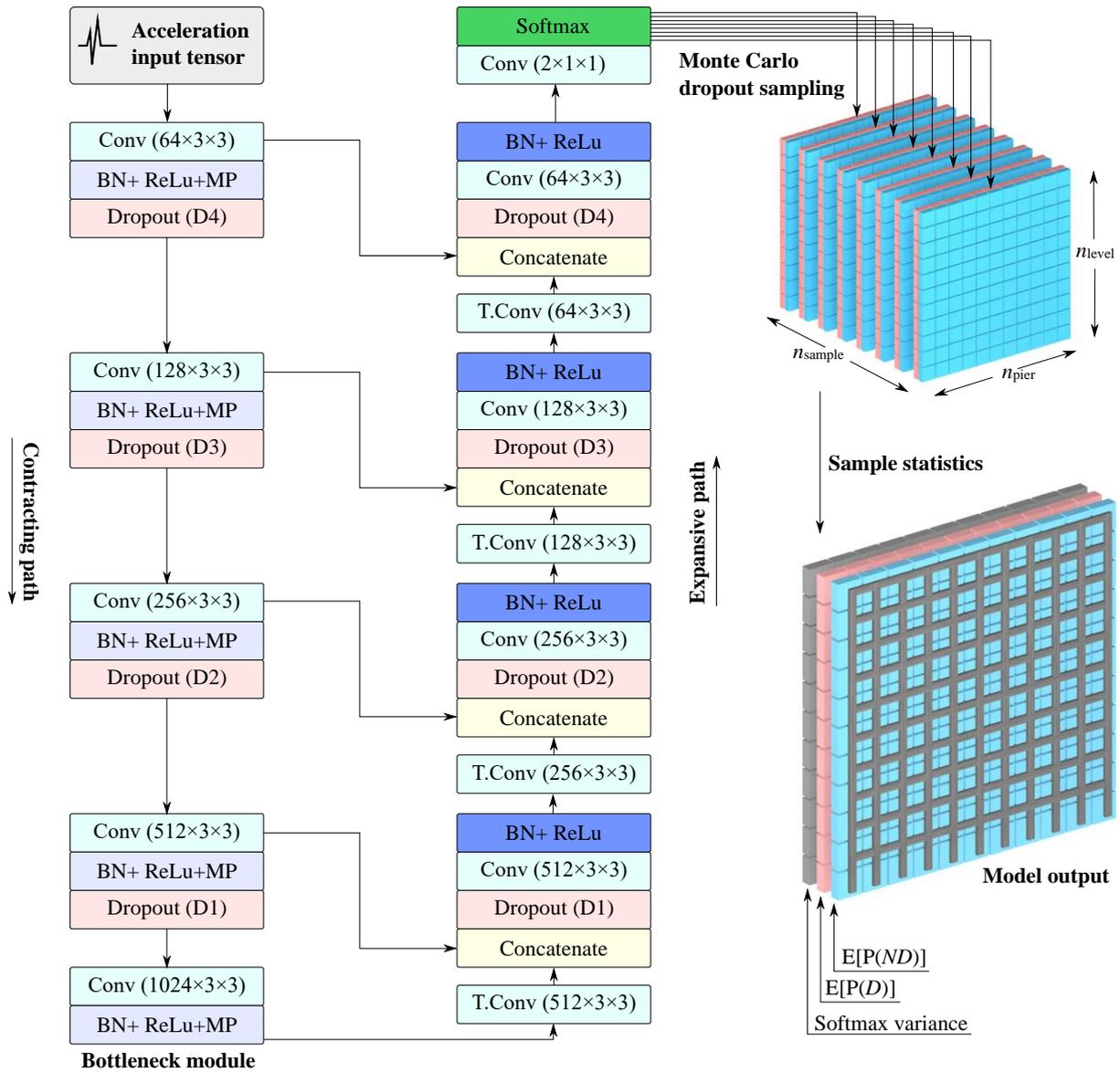

Fig. 2 – Bayesian U-Net for semantic damage segmentation. Conv (128×3×3): convolution layer that extracts 128 feature maps with a 3×3 kernel; BN: batch normalization; ReLu: rectified linear unit activation; MP: 2×2 max-pooling; T.Conv (256×3×3): transposed convolution that extracts 256 feature maps with a 3×3 kernel.

Dropout was initially proposed by Srivastava et al. [27] to alleviate overfitting in deep learning neural networks. Despite its regularization effects, Gal and Ghahramani [28] showed that the use of dropout at inference can be interpreted as a Bayesian approximation of the Gaussian process. As a result, the learnable parameters in a neural network and, subsequently, the prediction output of a deep learning model are random variables. This probabilistic model makes it possible to reason about the model uncertainty, which is of great value in damage predictions. The key difference between standard and Bayesian inference is that dropout operation is performed at test time. To approximate the distribution of output softmax probabilities, Monte Carlo dropout samples are drawn for each input observation (See Fig. 2).

The proposed architecture is designed to identify the location(s) of damage. Therefore, the softmax activation layer provides the binary output probabilities, including damage (P(*D*)) and no damage (ND) (P(*ND*)) for each structural node. Given the stochastic nature of the output, the expected sample values are used to predict the label of each node. Similarly, the variance of the sample is an appropriate metric to reason about the uncertainty. The binary class probabilities are mutually exclusive and collectively exhaustive. Therefore, the softmax variance is equal for both classes. Further discussion on the application of uncertainty is presented in Section 5.

## 4. Numerical Verification

In this section, a numerical case study is presented to investigate the performance and robustness of Bayesian U-Nets for damage segmentation. The dataset used in this regard is a 10-story-10-bay reinforced concrete moment frame as in [24], where the structure is simulated with 10,800 nonlinear response history analysis [29-31]. The same data splits of 0.8, 0.1, and 0.1 are respectively considered for training, validation, and testing of the models. Therefore, the results in this paper are comparable with the benchmark SDS-B model in [24]. In a grid node, exceeding the immediate occupancy limit defined in FEMA 365 [32] in any of the beams or column hinges is labeled as damage (*D*) in that location.

4.1. Dropout Hyperparameters

This section provides a parametric study on dropout hyperparameters and how they affect the overall performance of the classifier. We also discuss different strategies in selecting observation weights and decision rules to maximize the performance of Bayesian U-Net. Three types of hyperparameters can affect the performance of the model concerning dropout, including the probability of setting an input unit to zeros ($P_{do}$), the location of dropout layers, and finally, the number of Monte Carlo samples ($n_{sample}$). No dropout ($P_{do}=0$) increases the chances of overfitting and also means that the inference is not Bayesian. On the other hand, very high values of $P_{do}$ may result in underfitting and reduce the convergence speed of training. A bin of 13 possible $P_{do}$ values (0.02, 0.05, 0.08, 0.10, 0.12, 0.15, 0.20, 0.25, 0.30, 0.40, 0.50, 0.60, 0.70), and four different dropout layer configurations (DLCs) are considered, as shown in Table 1.

Table 1 – Dropout layer locations (consistent with the layer definition in Fig.2)

| Configuration | Dropout layer ID |
|---|---|
| DLC 1 | D1 |
| DLC 2 | D1, D2 |
| DLC 3 | D1, D2, D3 |
| DLC 4 | D1, D2, D3, D4 |

In this RC frame, 58% of observations correspond to the class of ND. Therefore, the model will tend to favor ND observations in training and minimizing the cross-entropy loss function. Since the correct prediction of damage (D) is more important than the ND class, the loss function can be balanced by the median frequency weight (MFW) assignment rather than considering uniform weights (UW). Other than modifying the loss function, adjusting the decision rule is another way of improving the damage class accuracy. The maximum likelihood (ML) and maximum-a-posteriori (MAP) are compared in assigning a label based on the expected class probabilities [33].

By considering all possible combinations, a total of 104 (13×4×2) models are independently trained and validated. In these initial investigations, 50 Monte Carlo dropout samples are generated for each observation. Later in section 4.2, the effects of sample size are studied. Nadam optimizer with a learning rate of 1.0E-4 and the exponential decay rate of 0.9996 per epoch is used to calibrate the learnable parameters of Bayesian U-Nets. Aside from dropout, L2 regularization with a factor of 1.0Ee-6 is also added to the loss function. All models are trained with a maximum number of 1000 epochs in mini-batches of 256 observations. An early stopping criterion of 200 epochs based on the validation loss is also used to terminate the training process when there is no improvement. It is noted that the validation loss is calculated based on Bayesian inference by taking the expected validation loss of Monte Carlo samples.

Table 2 – Summary of performance metrics for MAP decision rule

| | | Highest validation MCA | | | | |
|---|---|---|---|---|---|---|
| DLC | $P_{do}$ | Weight method | MCA (%) | GA (%) | $D$ class accuracy (%) | $ND$ class accuracy (%) |
| 4 | 0.40 | UW | 96.99 | 97.02 | 97.16 | 96.82 |
| 4 | 0.40 | MFW | 96.97 | 96.96 | 96.90 | 97.04 |
| 4 | 0.30 | MFW | 96.93 | 96.76 | 95.99 | 97.87 |
| 4 | 0.25 | MFW | 96.89 | 96.75 | 96.16 | 97.62 |
| 4 | 0.25 | UW | 96.86 | 96.69 | 95.97 | 97.75 |
| | | Lowest validation MCA | | | | |
| DLC | $P_{do}$ | Weight method | MCA (%) | GA (%) | $D$ class accuracy (%) | $ND$ class accuracy (%) |
| 4 | 0.70 | MFW | 94.45 | 94.40 | 94.18 | 94.72 |
| 4 | 0.70 | UW | 94.44 | 94.56 | 95.08 | 93.80 |
| 3 | 0.60 | MFW | 94.08 | 94.41 | 95.84 | 92.32 |
| 2 | 0.02 | UW | 91.49 | 90.96 | 88.66 | 94.32 |
| 1 | 0.05 | MFW | 89.19 | 89.38 | 90.20 | 88.18 |

Table 3– Summary of performance metrics for ML decision rule

| | | Highest validation MCA | | | | |
|---|---|---|---|---|---|---|
| DLC | $P_{do}$ | Weight method | MCA (%) | GA (%) | $D$ class accuracy (%) | $ND$ class accuracy (%) |
| 4 | 0.40 | UW | 96.65 | 96.58 | 96.26 | 97.04 |
| 4 | 0.40 | MFW | 96.60 | 96.48 | 95.96 | 97.24 |
| 4 | 0.25 | MFW | 96.60 | 96.41 | 95.59 | 97.61 |
| 4 | 0.25 | UW | 96.56 | 96.34 | 95.34 | 97.78 |
| 4 | 0.30 | MFW | 96.56 | 96.29 | 95.09 | 98.03 |
| | | Lowest validation MCA | | | | |
| DLC | $P_{do}$ | Weight method | MCA (%) | GA (%) | $D$ class accuracy (%) | $ND$ class accuracy (%) |
| 3 | 0.60 | MFW | 94.21 | 94.42 | 95.33 | 93.09 |
| 4 | 0.70 | UW | 93.62 | 93.39 | 92.36 | 94.88 |
| 4 | 0.70 | MFW | 93.61 | 93.25 | 91.68 | 95.54 |
| 2 | 0.02 | UW | 91.63 | 91.07 | 88.59 | 94.67 |
| 1 | 0.05 | MFW | 89.44 | 89.57 | 90.14 | 88.74 |

The backpropagation algorithm is performed for all epochs; however, the set of weights in the epoch with the lowest validation loss is selected for further evaluations. Global accuracy (*GA*) and mean class accuracy (*MCA*) are standard metrics where the latter is often preferred in imbalanced datasets as both class accuracies have equal weights. All models are sorted based on *MCA* results for the two decision rules of MAP and ML (Tables 2 and 3).

Placing the dropout layers before all computational blocks (DLC 4) yields the best model. In contrast, DLC 1 and 2 show inferior performance. Moreover, $P_{do}$=0.4 is the optimum dropping probability, while ratios between 0.25 and 0.5 seem to be appropriate considering validation *MCA* (see Fig. 3). It is observed that the models for relatively high $P_{do}$ do not converge within the 1000 epochs and may require longer training. In contrast, $P_{do}$ values of 2-5% increase the chances of overfitting. For further evaluation of the unseen test set, the model with the highest validation *MCA* (i.e., DLC 4 and $P_{do}$=0.4) is used.

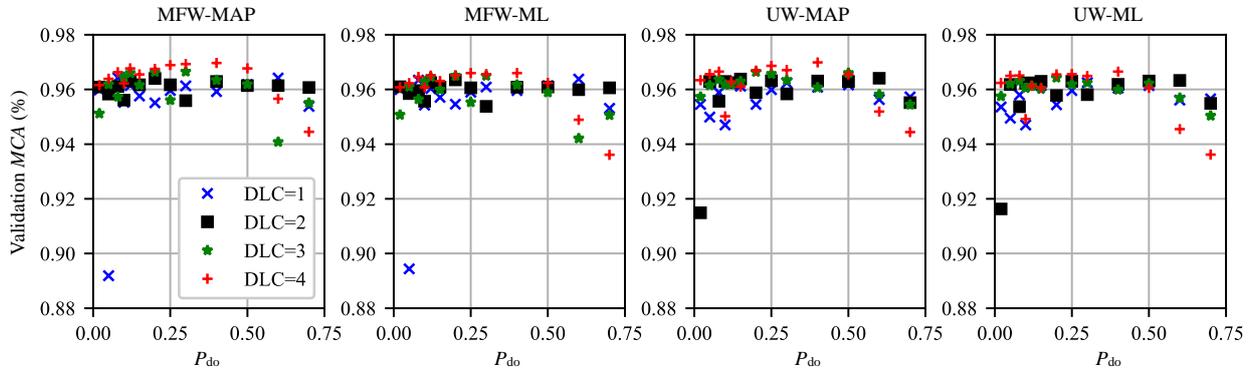

Fig. 3 – Effects of dropout configuration and ratio on *MCA*

4.2. Monte Carlo Sampling

The Bayesian inference is possible through generating a series of output samples that can be used to estimate the probability distribution of final predictions. An optimal number of samples is necessary to construct a reliable distribution while avoiding expensive costs of computation. In this section, the effect of sample size (number of Monte Carlo dropout samples) is investigated.

After the design of the dropout hyperparameters based on the validation set, the selected model is evaluated for unseen test data. Fig. 4 is a summary of the four performance metrics concerning different strategies and the number of dropout samples. Note that dropout is a random operator. Therefore, the generated output can be different, even assuming a constant value of $n_{sample}$. The same inference is repeated for ten independent trials to monitor the performance and stability of the Bayesian classifier. Each point in Fig.4a and 4b are, respectively, the average and standard deviation of the ten trials for the corresponding metric. It can be observed that $n_{sample}$>100 yields insignificant improvements accompanied with the increased cost of computations. Nevertheless, increasing the sample size improves the stability of the prediction results (lower standard deviation).

The training and evaluation of Bayesian U-Nets are conducted using an NVIDIA GTX 1080 GPU with 8 GB of memory. Keras API [34] is used to train and compile the deep learning architecture for Bayesian inference. The required time to feed an earthquake excitation and segment damage is presented in Table 4. The inference time is linearly increasing depending on the size of the Monte Carlo dropout sample. The results in this table verify the potential of deep Bayesian U-Nets for near real-time implementation. Thousands of such buildings can be evaluated in less than one minute assuming that the acceleration data is available on a cloud server.

Table 4 – Damage segmentation runtime of Bayesian U-Net

| $n_{sample}$ | 5 | 10 | 20 | 50 | 100 | 200 |
|---|---|---|---|---|---|---|
| Average inference time per observation (milliseconds) | 0.34 | 0.63 | 1.26 | 3.13 | 6.29 | 12.48 |

4.3. Robustness

In this section, a comparison of results between the Bayesian U-Net and the original SDS framework in [24] is presented. The datasets used in these evaluations are identical in every aspect. The original implementation in [20] provides two different testing sets denoted as ideal and stochastic data sets. While both are generated by considering 1,080 different earthquake simulations, the physical model is also different for each observation in the stochastic test set. The stochastic testing performance is more realistic because the variations in material properties and construction imperfections are considered in testing observations.

A summary of performance metrics is given in Table 5. The Bayesian U-Net is superior in all strategies when compared with the benchmark SDS. While all strategies are close in performance, the UW-MAP is more accurate considering *GA* and *MCA*. However, MFW-MAP and MFW-ML show higher accuracies for the damaged class (Fig.4a). A decision-maker will probably favor such strategies as higher prediction accuracy of damage class is commonly preferred over minimizing false alarms.

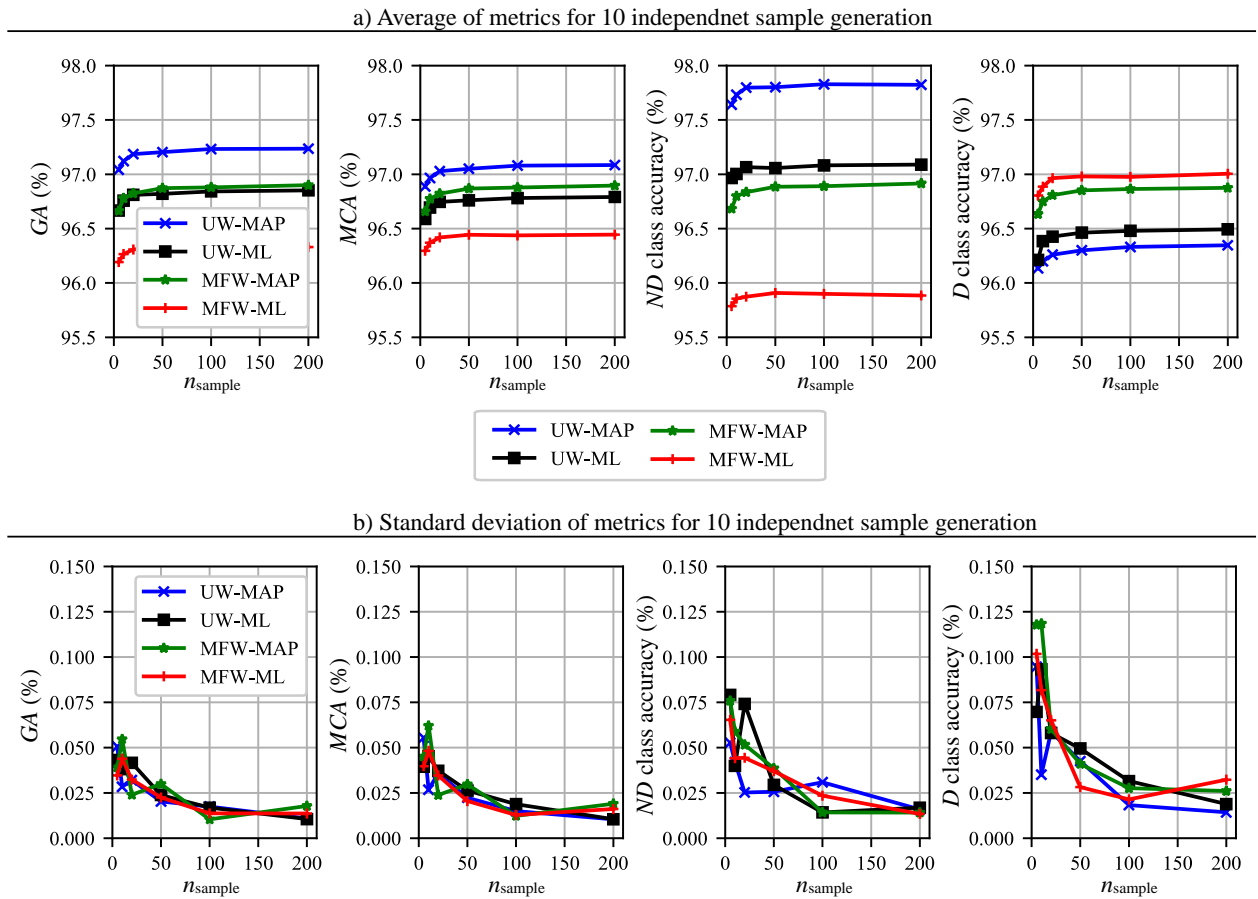

Fig. 4 – Effects of the number of samples on the robustness of Bayesian U-Net

## 5. Uncertainty Quantification

Up to this point, we have shown that Bayesian U-Net is a robust classifier that can be used to localize damage. However, the chances of misclassifications, no matter how small, still exist. For a safe implementation of the SDS framework, the consideration of the model uncertainty is crucial. As it is shown in Fig. 2, the output of the model includes the softmax variance and the expected class probabilities. Note that in binary damage localization, the variance is equal for both classes (despite the difference in expected).

The softmax variance is an integral part of the output that is readily available at the test time without requiring additional input. The model uncertainty is a two-dimensional mask which is stored as an extra channel in the prediction output tensor. The normalized softmax variance mask (normalized by the highest uncertainty) for several selected observations is shown in Fig 5. It can be observed that the locations of the misclassified structural nodes mostly correlate well with regions of high model uncertainty (darker regions). The absolute value of model uncertainty also aides a decision-maker to better interpret the normalized softmax variances. For example, the last observation corresponds to no damage, and the prediction is 100% accurate. Compared to other observations, the maximum softmax variance of this example is relatively small (0.08%), showing its confidence in the predictions.

Table 5 – Test set performance of Bayesian U-Net compared with benchmark models

| | Ideal data set | | | |
|---|---|---|---|---|
| | *MCA* (%) | | | |
| | UW-MAP | UW-ML | MFW-MAP | MFW-ML |
| benchmark SDS | 95.80 | 95.88 | 96.14 | 95.86 |
| Bayesian U-Net | **97.08** | **96.82** | **96.85** | **96.45** |
| | *GA* (%) | | | |
| | UW-MAP | UW-ML | MFW-MAP | MFW-ML |
| benchmark SDS | 96.09 | 95.78 | 96.13 | 96.09 |
| Bayesian U-Net | **97.22** | **96.88** | **96.80** | **96.34** |
| | Stochastic data set | | | |
| | *MCA* (%) | | | |
| | UW-MAP | UW-ML | MFW-MAP | MFW-ML |
| benchmark SDS | 94.55 | 94.70 | 94.80 | 94.31 |
| Bayesian U-Net | **95.41** | **95.22** | **95.33** | **94.99** |
| | *GA* (%) | | | |
| | UW-MAP | UW-ML | MFW-MAP | MFW-ML |
| benchmark SDS | 94.89 | 94.64 | 94.84 | 94.65 |
| Bayesian U-Net | **95.61** | **95.34** | **95.39** | **94.91** |

## 6. Conclusions

Deep learning portrays a promising application for the future of SHM. Real-time post-disaster inspections can be significantly accelerated, utilizing state-of-the-art AI algorithms. Nonetheless, certain challenges remain for the reliable implementation of data-driven models in SHM. This paper proposes the Bayesian U-Nets for semantic damage segmentation (SDS) of large scale structures. Deep Bayesian neural networks provide an uncertainty output, which is crucial for a decision-maker who needs to take action based on safety assessments.

In this paper, the prediction uncertainty of U-Nets is estimated by activating dropout at the inference time. Dropout hyperparameters, including the layer configuration and probability of this operator, are investigated. Furthermore, several strategies in modifying the observation weights in the loss function and adopting different decision rules are compared on different performance metrics. The configuration with a dropout layer after each computational block results in the models with the highest *GA* and *MCA*. Moreover, dropout probabilities in the range of 0.25 to 0.5 are shown to be appropriate for the designed architecture in this case study.

After tuning the network hyperparameters on the validation set, the runtime and stability of the algorithm are evaluated with different sample sizes. It is shown that even for the sample size as large as 200, the classifier is still capable of performing near real-time damage diagnosis for post-disaster inspections. Compared to the benchmark classifier in [24], the Bayesian U-Net achieves superior testing performance in terms of all four metrics.

The proposed framework boosts *MCA* and *GA* to values above 95% and 97% accuracy respectively for the ideal and stochastic test sets. Despite a relatively high accuracy, the risk of misclassifications still exits. To implement SDS more reliable, the softmax variance is shown to be very effective. Several examples are provided to show that when the model uncertainty is high, chances risks of prediction mistakes increases. Knowing the model uncertainty will help the users of such a data-driven model to make informed decisions in their future planning.

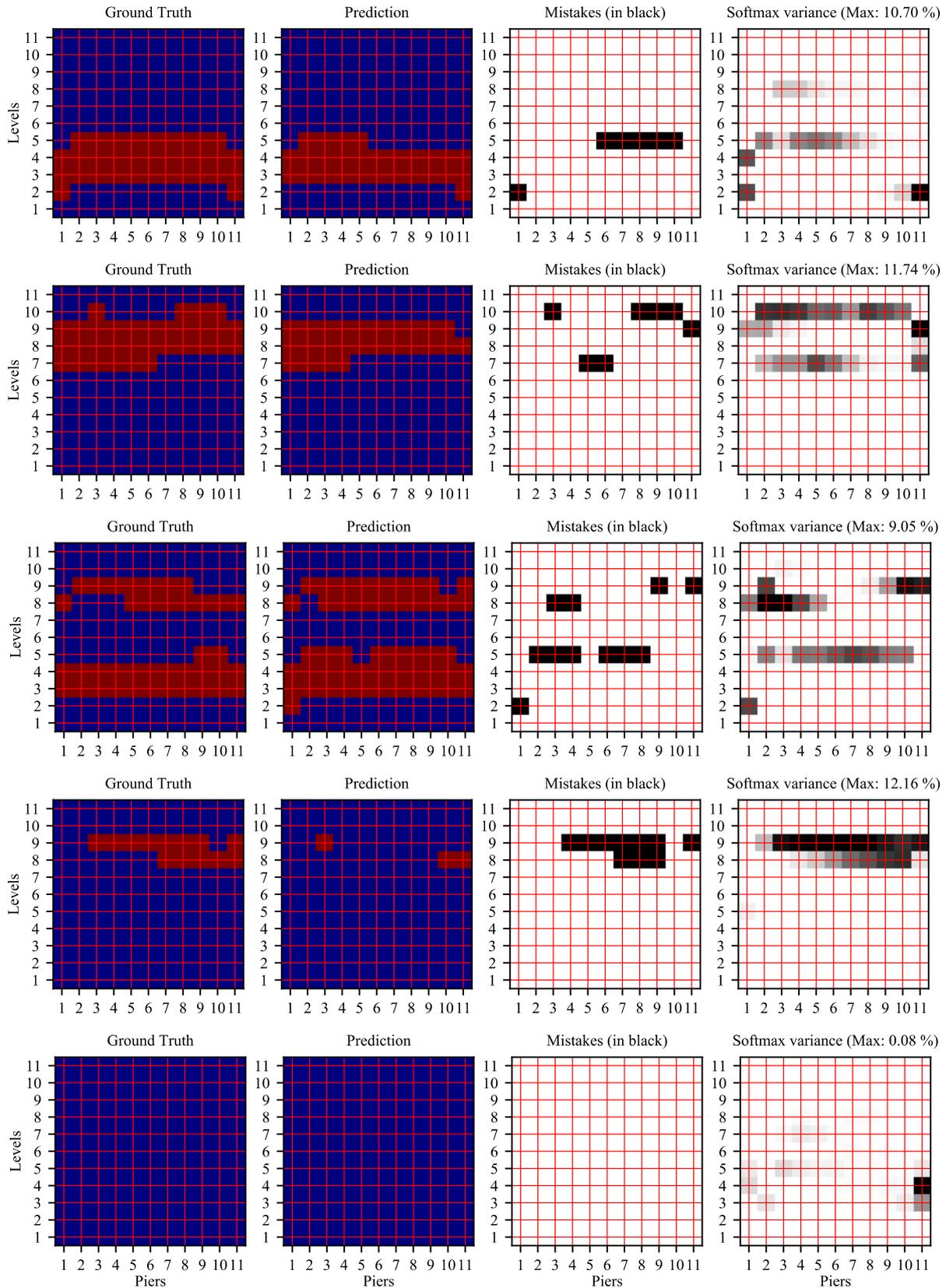

Fig. 5 – Correlations between mistakes and the model's uncertainty output. The nodes highlighted in red (in the ground truth and prediction masks) denote the presence of damage.